\documentclass[pra,twocolumn,amsmath,amssymb]{revtex4}
\usepackage{graphicx}
\usepackage{bm,fancybox}
\usepackage{amsmath,amssymb}
\usepackage{color}
\usepackage{ascmac}

\begin{document}
\title{The process matrix framework for
a single-party system}
\author{Tomoyuki Morimae}
\email{morimae@gunma-u.ac.jp}
\affiliation{ASRLD Unit, Gunma University,
1-5-1 Tenjin-cho Kiryu-shi Gunma-ken, 376-0052, Japan}
\date{\today}
            
\begin{abstract}
The process matrix framework
[O. Oreshkov, F. Costa, and C. Brukner, Nature Communications {\bf3}, 1092 (2012)]
can describe general physical theory where locally
operations are described by completely-positive maps
but globally no fixed causal structure is assumed.
In this framework, two parties who perform measurements on each single-qubit system
can violate a ``causal inequality",
which is not violated if the global fixed causal structure exists.
Since the standard quantum physics assumes a fixed global causal structure,
the process matrix framework can describe more general physical theory than the
standard quantum physics.
In this paper, we show that 
for a single-party system  
the process matrix framework 
is reduced to the standard quantum physics,
and therefore no exotic effect beyond the standard quantum 
physics can be observed.
This result is analogous to the well known fact in the Bell inequality violation:
a single-party system can be described by a local hidden variable theory,
whereas more than two parties can violate the Bell inequality.
\end{abstract}

\maketitle  

\section{Introduction}
Exploring more general physical theory beyond the standard quantum physics
has great practical importance as well as pure academic interest.
It can give some (and hopefully full)
explanations why quantum physics is as it is while
quantum physics is not the most general no-signaling theory~\cite{PR,IC,Gross,Barnum,Torre}.
It also provides plenty of new insights for researches in
other fields,
such as statistical physics, field theory, and, interestingly, even computer science.
For example, it is known that certain exotic effects, such as
the closed-time-like curve~\cite{Bacon,AaronsonCTC}, nonlinear time evolutions~\cite{Abrams}, 
and postselections~\cite{AaronsonpostBQP}, enable 
super strong computing (such as PP). It was also pointed out that
the fact that universal quantum computing with
postselections (postBQP) is very strong (i.e., PP)~\cite{AaronsonpostBQP} can be used to
show the hardness of
classical efficient simulations of some superficially innocent 
non-universal quantum computing models,
such as depth-four circuits~\cite{Terhal}, 
commuting gates~\cite{IQP}, 
non-interacting bosons~\cite{Bosonsampling},
and the one-clean qubit model (DQC1)~\cite{KL,MFF}.

There are several theoretical formalisms to study general physics
beyond the standard quantum physics~\cite{Oreshkov,Pavia,Peri,Joe}.
Oreshkov, Costa, and Brukner~\cite{Oreshkov}
recently proposed a new framework, so called the process matrix (PM) framework, to study 
a physical theory where locally 
operations are described by CP maps but
globally no fixed causal structure
is assumed (see also Refs.~\cite{Brukner1,Brukner2,SWolf,SWolf2,MorimaePM,Hardy1,Hardy2,Hardy3}). 
Since the standard quantum physics assumes the global fixed causal structure,
the PM framework can describe more general physical theory beyond
the standard quantum physics.
In fact, it was shown in Ref.~\cite{Oreshkov} that 
for two parties who perform measurements on each single-qubit system
the PM framework can violate a ``causal inequality",
which is not violated in the theory where the global fixed causal
structure exists. 
(See the next section.)

Does the PM framework always exhibit some exotic effects beyond 
the standard quantum physics?
Or exotic effects are exceptional for some special circumstances, such as
certain specific system dimension or number of parties, etc.?

The purpose of the present paper is to study the question.
In this paper, 
we show that for 
a single-party system
we cannot see any exotic effect beyond the standard quantum physics:
the PM framework is reduced to the standard quantum physics.
It is interesting to point out that
this result has an analogy in the Bell inequality violation:
a single-party system can be described by a local hidden variable theory,
whereas more than two parties can 
violate the Bell inequality (or its multipartite generalizations).

\section{Two parties experiment}
\begin{figure}[htbp]
\begin{center}
\includegraphics[width=0.4\textwidth]{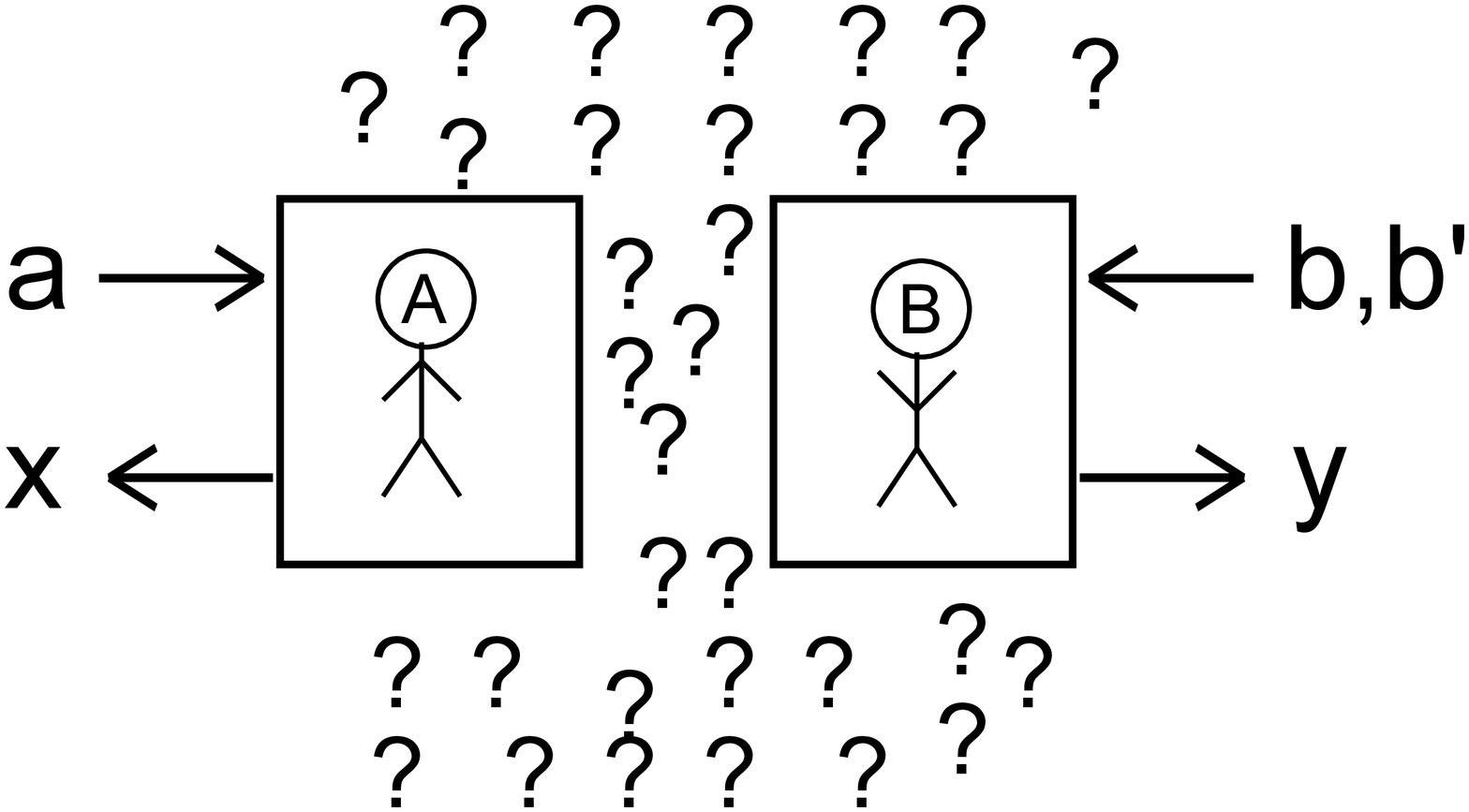}
\end{center}
\caption{
The causal game considered in Ref.~\cite{Oreshkov}.
Alice and Bob are in different laboratories.
Operations in the inside of each laboratory is described by
CP maps,
but in the outside of the laboratories
no fixed global causal structure is assumed.
} 
\label{game}
\end{figure}

It was shown in Ref.~\cite{Oreshkov}
that the PM framework for two parties
exhibits the exotic effect, namely, the violation
of the causal inequality.
Let us consider the following game (Fig.~\ref{game}).
Alice and Bob are in the different laboratories.
Operations in the inside of each laboratory are described
by CP maps,
but in the outside of the laboratories, no fixed causal structure is assumed. 
Alice is given a random bit $a\in\{0,1\}$ and has to output $x\in\{0,1\}$.
Bob is given two random bits $b,b'\in\{0,1\}$ and has to output $y\in\{0,1\}$.
A quantum state enters into Alice's laboratory, and she performs a measurement
on it. She sends the post-measurement state out to the laboratory.
Also in Bob's laboratory, he performs a measurement on
an entering quantum state, and sends the post-measurement state out
to his laboratory.

If a fixed global causal structure exists
in the outside of the laboratories,
the inequality
\begin{eqnarray*}
p_{OCB}\equiv\frac{1}{2}\Big[
p(x=b|b'=0)
+p(y=a|b'=1)
\Big]
\le\frac{3}{4}
\end{eqnarray*}
is satisfied~\cite{Oreshkov}.
This bound is called the causal inequality.
(Analogous to the Bell inequality, which is satisfied if the
system is described by a local hidden variable theory.)
They showed that in the PM framework,
this inequality can be violated.
Assume that Alice's measurement is a CP map 
\begin{eqnarray*}
{\mathcal E}_j^A:L(H_1^A)\to L(H_2^A),
\end{eqnarray*}
and Bob's measurement is a CP map 
\begin{eqnarray*}
{\mathcal E}_k^B:L(H_1^B)\to L(H_2^B),
\end{eqnarray*}
where $H_1^A$ and $H_2^A$ are Alice's input and output Hilbert spaces,
respectively,
$H_1^B$ and $H_2^B$ are Bob's input and output Hilbert spaces,
respectively,
and  $L(H)$ is the space of operators over $H$.
In the PM framework, the probability of having this event is given by
\begin{eqnarray*}
P({\mathcal E}_j^A,{\mathcal E}_k^B)=\mbox{Tr}
\Big[W\Big(M_{{\mathcal E}_j^A}\otimes M_{{\mathcal E}_k^B}\Big)\Big],
\end{eqnarray*}
where 
\begin{eqnarray*}
W\in L(H_1^A\otimes H_2^A\otimes H_1^B\otimes H_2^B)
\end{eqnarray*}
is a positive-semidefinite operator so-called the process matrix (PM).
A PM is considered as
a generalization of a density matrix in the standard quantum theory.
It was shown in Ref.~\cite{Oreshkov} that
$W$ has to be positive semi-definite, 
\begin{eqnarray*}
W\ge0.
\end{eqnarray*}
Operators
\begin{eqnarray*}
M_{{\mathcal E}_j^A}&\in& L(H_1^A\otimes H_2^A),\\ 
M_{{\mathcal E}_k^B}&\in& L(H_1^B\otimes H_2^B),
\end{eqnarray*}
are Choi-Jamiolkowsky (CJ) 
operators corresponding to
${\mathcal E}_j^A$
and ${\mathcal E}_k^B$, respectively.
Here, the CJ operator $M_{{\mathcal E}_j}$ corresponding to a CP map
\begin{eqnarray*}
{\mathcal E}_j:L(H_1)\to L(H_2)
\end{eqnarray*}
is defined by
\begin{eqnarray*}
M_{{\mathcal E}_j}&\equiv&[(I\otimes {\mathcal E}_j)|ME\rangle\langle ME|]^T\\
&=&\sum_{i,j=1}^{d_1}|i\rangle\langle j|\otimes{\mathcal E}_j(|j\rangle\langle i|)
\in L(H_1\otimes H_2),
\end{eqnarray*}
where $d_1$ is the dimension of $H_1$,
$T$ is the matrix transposition,
and 
\begin{eqnarray*}
|ME\rangle\equiv\sum_{j=1}^{d_1}|j\rangle\otimes|j\rangle\in
H_1\otimes H_1
\end{eqnarray*}
is the (non-normalized) maximally-entangled state.

For example, if the CP operation is to project onto $|\psi\rangle$
and change the post-measurement state into $|\eta\rangle$,
the corresponding CJ operator is
\begin{eqnarray*}
\sum_{i,j=1}^{d_1}|i\rangle\langle j|\otimes
|\eta\rangle\langle\psi|j\rangle\langle i|\psi\rangle\langle\eta|
=
\ovalbox{$\psi$}\otimes\ovalbox{$\eta$},
\end{eqnarray*}
where $\ovalbox{$x$}\equiv|x\rangle\langle x|$~\cite{Vedrans}.

In order to see a violation of the causal inequality,
authors of Ref.~\cite{Oreshkov} proposed the following PM:
\begin{eqnarray*}
W_{OCB}\equiv\frac{1}{4}\Big[
I^{\otimes 4}
+
\frac{
I\otimes Z\otimes Z\otimes I
+Z\otimes I\otimes X\otimes Z
}{\sqrt{2}}
\Big].
\end{eqnarray*}
If Alice's CJ operator is 
\begin{eqnarray*}
\ovalbox{$x$}\otimes\ovalbox{$a$},
\end{eqnarray*}
and Bob's CJ operator is
\begin{eqnarray*}
b'\ovalbox{$y$}\otimes \ovalbox{$\eta$}
+(b'\oplus1)\ovalbox{$(-1)^y$}\otimes \ovalbox{$b+y$},
\end{eqnarray*}
where $|\eta\rangle$ is any state,
then,
\begin{eqnarray*}
p_{OCB}
=\frac{2+\sqrt{2}}{4}>\frac{3}{4}.
\end{eqnarray*}
In this way, the PM framework for two laboratories
can exhibit the exotic effect beyond the standard quantum physics.

\section{Single-party system}
\begin{figure}[htbp]
\begin{center}
\includegraphics[width=0.45\textwidth]{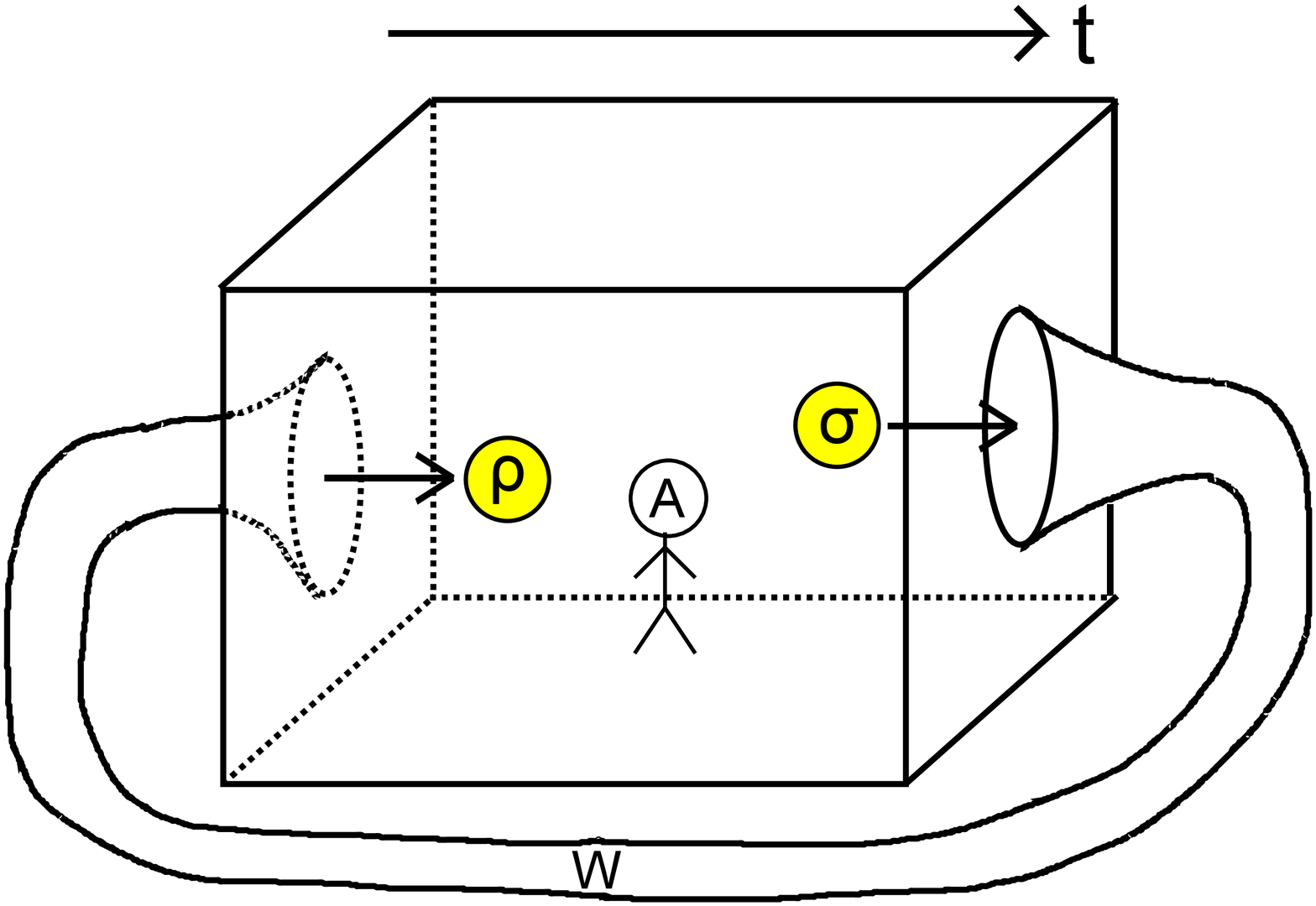}
\end{center}
\caption{Alice is in her laboratory. She performs a measurement on 
the state $\rho$ entered into the laboratory, and sends the post-measurement
state $\sigma$ out to the laboratory.
Her operation in the inside of the laboratory is described by a CP map, but
there is no fixed causal structure in the outside of the laboratory.} 
\label{single}
\end{figure}

Now let us show that the PM framework is reduced to the
standard quantum physics for a single-party system.
We assume that Alice is in her laboratory (Fig.~\ref{single}).
A state $\rho$ enters into the laboratory, and she measures
it. Then she sends the post-measurement state $\sigma$ out
to the laboratory.
Her operation in the inside of the laboratory is described by
a CP map,
but there is no fixed causal structure in the outside of her laboratory.

In the PM framework, this experiment is described
in the following way.
Alice's measurement on the entering state $\rho$
is described by a
CP map
\begin{eqnarray*}
{\mathcal E}_j:L(H_1)\to L(H_2)
\end{eqnarray*}
if the measurement result is $j$,
where $H_1$ and $H_2$ 
are the input and output Hilbert spaces, respectively.
The probability 
$P({\mathcal E}_j)$ that 
Alice's measurement result is $j$ is given by
\begin{eqnarray*}
P({\mathcal E}_j)=\mbox{Tr}(W M_{{\mathcal E}_j}),
\end{eqnarray*}
where 
\begin{eqnarray*}
M_{{\mathcal E}_j}&\equiv&[(I\otimes {\mathcal E}_j)|ME\rangle\langle ME|]^T\\
&=&\sum_{i,j=1}^{d_1}|i\rangle\langle j|\otimes{\mathcal E}_j(|j\rangle\langle i|)
\in L(H_1\otimes H_2)
\end{eqnarray*}
is the CJ operator
corresponding to ${\mathcal E}_j$, and
the operator
\begin{eqnarray*}
W\in L(H_1\otimes H_2), 
\end{eqnarray*}
is the PM.
It is easy to show 
that
\begin{eqnarray*}
\mbox{Tr}(W)=d_2,
\end{eqnarray*}
where $d_2$ is the dimension of $H_2$
(for a proof, see Appendix).
In other words,
the PM $W$ is a density matrix up to the constant factor
$d_2$. 

For example, if Alice's CJ operator is 
$\ovalbox{$\psi$}\otimes\ovalbox{$\eta$}$,
the probability for the event  
is given by
\begin{eqnarray*}
\mbox{Tr}\Big[W\Big(\ovalbox{$\psi$}\otimes\ovalbox{$\eta$}\Big)\Big].
\end{eqnarray*}
One might think that if 
$W$ could be a maximally-entangled state between $H_1$ and $H_2$,
then the closed time-like curve can be implemented, since
Alice can ``postselect" the $H_2$ part of $W$ into $|\eta\rangle$,
which affects the ``past state", i.e., the $H_1$ part of $W$.

However, such a ``strange" effect does not happen.
We now show that in the single-party setup
the PM framework is reduced to the standard quantum physics.
(For simplicity, we here give a proof for the
single-qubit case. A generalization to the multi-qubit case is given in Appendix.)
Let us decompose $W$ in the Pauli basis: 
\begin{eqnarray*}
W=\sum_{\alpha\in\{1,x,y,z\}}
\sum_{\beta\in\{1,x,y,z\}}
w_{\alpha\beta}\sigma_\alpha\otimes\sigma_\beta,
\end{eqnarray*}
where $\{w_{\alpha\beta}\}$ are complex numbers, and
\begin{eqnarray*}
\sigma_1&\equiv& I=|0\rangle\langle0|+|1\rangle\langle1|,\\
\sigma_x&\equiv& X=|0\rangle\langle1|+|1\rangle\langle0|,\\
\sigma_y&\equiv& Y=-i|0\rangle\langle1|+i|1\rangle\langle0|,\\
\sigma_z&\equiv& Z=|0\rangle\langle0|-|1\rangle\langle1|,
\end{eqnarray*}
are Pauli operators.

Since $\mbox{Tr}(W)=2$, we obtain $w_{11}=\frac{1}{2}$.
For $\alpha\in\{x,y,z\}$
and $\beta\in\{x,y,z\}$,
let $|\alpha(s)\rangle$ be the eigenvector of $\sigma_\alpha$
corresponding to the eivenvalue $(-1)^s$ $(s\in\{0,1\})$,
and $|\beta(m)\rangle$ be the eigenvector of $\sigma_\beta$
corresponding to the eivenvalue $(-1)^m$ $(m\in\{0,1\})$.
\begin{eqnarray*}
1&=&\sum_{s=0}^1\mbox{Tr}\Big[W
\Big(
\ovalbox{$\alpha(s)$}\otimes\ovalbox{$\beta(m)$}\Big)
\Big]\\
&=&\sum_{s=0}^1
\Big(
w_{1,1}
+(-1)^m w_{1,\beta}
+(-1)^s w_{\alpha,1}
+(-1)^{s+m}w_{\alpha,\beta}\Big)\\
&=&1+
\sum_{s=0}^1
\Big(
(-1)^m w_{1,\beta}
+(-1)^{s+m}w_{\alpha,\beta}\Big),
\end{eqnarray*}
which leads to
\begin{eqnarray*}
\sum_{s=0}^1
\Big(
(-1)^m w_{1,\beta}
+(-1)^{s+m}w_{\alpha,\beta}\Big)=0.
\end{eqnarray*}
If we take $m=s$, then
\begin{eqnarray*}
0=
\sum_{s=0}^1
\Big(
(-1)^s w_{1,\beta}
+(-1)^{2s}w_{\alpha,\beta}\Big)
=
2w_{\alpha,\beta}.
\end{eqnarray*}
If we take $m=0$, then
\begin{eqnarray*}
0=
\sum_{s=0}^1
\Big(
(-1)^0 w_{1,\beta}
+(-1)^sw_{\alpha,\beta}\Big)
=
2w_{1,\beta}.
\end{eqnarray*}

In summary, we have shown that
for the single-qubit system the PM has the form of
\begin{eqnarray*}
W=\Big(\frac{1}{2}I
+w_{x1}X
+w_{y1}Y
+w_{z1}Z
\Big)\otimes I
\equiv W_1\otimes I.
\end{eqnarray*}
Since $W\ge0$, we obtain $W_1\ge0$.
This fact and $\mbox{Tr}(W_1)=1$ means that
$W_1$ is a density operator.
In this case, the PM framework recovers the standard quantum physics,
since for any Kraus operator $E_k$,
\begin{eqnarray*}
&&\mbox{Tr}\Big[W\Big(
\sum_{i,j}|i\rangle\langle j|\otimes 
\sum_k
E_k|j\rangle\langle i|E_k^\dagger
\Big)\Big]\\
&=&
\mbox{Tr}\Big[(W_1\otimes I)\Big(
\sum_{i,j}|i\rangle\langle j|\otimes 
\sum_k
E_k|j\rangle\langle i|E_k^\dagger
\Big)\Big]\\
&=&
\sum_{i,j,k}
\langle j|W_1|i\rangle
\langle i|E_k^\dagger E_k|j\rangle\\
&=&
\mbox{Tr}\Big(\sum_k E_k W_1 E_k^\dagger\Big),
\end{eqnarray*}
which is the probability rule in
the standard quantum physics.

\section{Discussion}
In this paper, we have shown that the process matrix 
framework for a single-party system is reduced to the
standard quantum physics. This result has the analogy in the Bell
inequality violation. It will be a future research subject
to clarify whether
the analogy is only a superficial one
or some deep connections are underlying.
For example, the causal inequality
has the same bound as that of the Bell inequality.
Furthermore, an upperbound of the causal inequality was derived~\cite{Brukner1},
which is an analog to the Tsirelson bound of the Bell inequality. 

One consequence of our result for experiments would be the following:
imagine that an experimental project team is trying to test quantum physics.
In order to capture any exotic effect beyond the standard quantum physics,
like a closed time like curve,
they have constructed a large 
accelerator by spending huge amount of budget.
However, due to the shortage of money, they have managed to
get only a single laboratory.
Can they see any exotic effect beyond the standard quantum physics?
Our result shows that
unless the ``CP map description rule" is locally wrong
they cannot see any exotic effect with a single laboratory.
In other words, they have to find new sponsor to construct another laboratory.

\acknowledgements
This work was supported by the Tenure Track System MEXT Japan
and the KAKENHI 26730003 by JSPS.

{\bf Note added}.---
After uploading (the first version of) this paper on arXiv, the author was informed that
the result was already shown in Refs.~\cite{Giulio_note,Giulio_note2}
and also given in the supplementary material of
Ref.~\cite{Oreshkov}.

\appendix*
\section{}
We first show $\mbox{Tr}(W)=d_2$.
Let us consider the Kraus operator
\begin{eqnarray*}
E_{j,k}\equiv \frac{1}{\sqrt{d_2}}|j\rangle\langle k|
\end{eqnarray*}
for $j=1,2,...,d_2$ and
$k=1,2,...,d_1$,
which satisfies
\begin{eqnarray*}
\sum_{j=1}^{d_2}\sum_{k=1}^{d_1}E_{j,k}^\dagger E_{j,k}
&=&\frac{1}{d_2}\sum_{j=1}^{d_2}\sum_{k=1}^{d_1}
|k\rangle\langle j|j\rangle\langle k|\\
&=&I_{d_1}.
\end{eqnarray*}
Then, we obtain
\begin{eqnarray*}
1
&=&\sum_{j=1}^{d_2}\sum_{k=1}^{d_1}
\mbox{Tr}\Big(WM_{E_{j,k}}\Big)\\
&=&\frac{1}{d_2}\sum_{j=1}^{d_2}\sum_{k=1}^{d_1}
\mbox{Tr}\Big[W\Big(\ovalbox{$k$}\otimes \ovalbox{$j$}\Big)\Big]\\
&=&\frac{1}{d_2}\mbox{Tr}(W),
\end{eqnarray*}
which means $\mbox{Tr}(W)=d_2$.

We next show the generalization of the single-qubit result
to multi-qubit result.
Let $S_{s_1,...,s_n}$ be a certain subset of $\{0,1\}^n$,
which can depend on $(s_1,...,s_n)\in\{0,1\}^n$.
\begin{widetext}
\begin{eqnarray*}
1&=&
\sum_{(s_1,...,s_n)\in\{0,1\}^n}
\frac{1}{|S_{s_1,...,s_n}|}
\sum_{(m_1,...,m_n)\in S_{s_1,...,s_n}}
\mbox{Tr}\Big(W
\ovalbox{$\alpha_1(s_1)$}
\otimes...\otimes
\ovalbox{$\alpha_n(s_n)$}
\otimes
\ovalbox{$\beta_1(m_1)$}
\otimes...\otimes
\ovalbox{$\beta_n(m_n)$}
\Big)\\
&=&
\sum_{(s_1,...,s_n)\in\{0,1\}^n}
\frac{1}{|S_{s_1,...,s_n}|}
\sum_{(m_1,...,m_n)\in S_{s_1,...,s_n}}
\sum_{\xi_1\in\{1,\alpha_1\}}
...
\sum_{\xi_n\in\{1,\alpha_n\}}
\sum_{\eta_1\in\{1,\beta_1\}}
...
\sum_{\eta_n\in\{1,\beta_n\}}\\
&&\times w_{\xi_1,...,\xi_n,\eta_1,...,\eta_n}
\mbox{Tr}\Big(
\sigma_{\xi_1}
\ovalbox{$\alpha_1(s_1)$}
\Big)
...
\mbox{Tr}\Big(
\sigma_{\xi_n}
\ovalbox{$\alpha_n(s_n)$}
\Big)
\mbox{Tr}\Big(
\sigma_{\eta_1}
\ovalbox{$\beta_1(m_1)$}
\Big)
...
\mbox{Tr}\Big(
\sigma_{\eta_n}
\ovalbox{$\beta_n(m_n)$}
\Big).
\end{eqnarray*}
\end{widetext}

Consider certain $w_{1,...,1,\eta_1,...,\eta_n}$,
where  
$\eta_j$,$\eta_k$,...,$\eta_r$ are not 1.
If we take 
\begin{eqnarray*}
S_{s_1,...,s_n}=\Big\{(m_1,...,m_n)\Big|
m_j\oplus m_k\oplus...\oplus m_r=0\Big\},
\end{eqnarray*}
then
\begin{eqnarray*}
1=2^nw_{1,...,1,1,...,1}+2^nw_{1,...,1,\eta_1,...,\eta_n},
\end{eqnarray*}
which means that
$
w_{1,...,1,\eta_1,...,\eta_n}=0.
$

Next consider certain $w_{\xi_1,...,\xi_n,\eta_1,...,\eta}$,
where $\xi_t,\xi_u,...,\xi_v$ and $\eta_j,\eta_k,...,\eta_r$
are not 1.
Then, if we take 
\begin{eqnarray*}
&&S_{s_1,s_2,...,s_n}\\
&=&\Big\{(m_1,...,m_n)\Big|
s_t\oplus s_u\oplus...\oplus s_v=m_j\oplus m_k\oplus...\oplus m_r\Big\},
\end{eqnarray*}
then
\begin{eqnarray*}
1=2^nw_{1,...,1,1,...,1}+2^nw_{\xi_1,...,\xi_n,\eta_1,...,\eta_n},
\end{eqnarray*}
which means that
$
w_{\xi_1,...,\xi_n,\eta_1,...,\eta_n}=0.
$



\begin{thebibliography}{99}
\bibitem{PR}
S. Popescu and D. Rohrlich,
Found. Phys. {\bf24}, 379 (1994).

\bibitem{IC}
M. Pawlowski, T. Paterek, D. Kaszlikowski, V. Scarani, A. Winter,
and M. Zukowski,
Nature {\bf461}, 1101 (2009).

\bibitem{Gross}
D. Gross, M. Mueller, R. Colbeck, and O. C. O. Dahlsten,
Phys. Rev. Lett. {\bf104}, 080402 (2010).

\bibitem{Barnum}
H. Barnum, S. Beigi, S. Boixo, M. B. Elliott, and S. Wehner,
Phys. Rev. Lett. {\bf104}, 140401 (2010).

\bibitem{Torre}
G. de la Torre, L. Masanes, A. J. Short, and M. P. Mueller,
Phys. Rev. Lett. {\bf109}, 090403 (2012).

\bibitem{Bacon}
D. Bacon,
Phys. Rev. A {\bf70}, 032309 (2004).

\bibitem{AaronsonCTC}
S. Aaronson and J. Watrous,
Proc. R. Soc. A {\bf465}, 631 (2009).

\bibitem{Abrams}
D. S. Abrams and S. Lloyd,
Phys. Rev. Lett. {\bf81}, 3992 (1998).

\bibitem{AaronsonpostBQP}
S. Aaronson, Proc. R. Soc. A {\bf461}, 3437 (2005).

\bibitem{Terhal}
B. Terhal and D. DiVincenzo, Quant. Inf. Comput. {\bf4}, 134 (2004).

\bibitem{IQP}
M. J. Bremner, R. Jozsa, and D. J. Shepherd,
Proc. R. Soc. A {\bf467}, 2126 (2011).

\bibitem{Bosonsampling}
S. Aaronson and A. Arkhipov,
Theory of Computing {\bf9}, 143 (2013).

\bibitem{KL}
E. Knill and R. Laflamme,
Phys. Rev. Lett. {\bf81}, 5672 (1998).

\bibitem{MFF}
T. Morimae, K. Fujii, and J. F. Fitzsimons,
Phys. Rev. Lett. {\bf112}, 130502 (2014).

\bibitem{Pavia}
G. Chiribella, G. M. D'Ariano, P. Perinotti, and B. Valiron,
Phys. Rev. A {\bf88}, 022318 (2013).

\bibitem{Peri}
M. S. Leifer and R. W. Spekkens,
Phys. Rev. A {\bf88}, 052130 (2013).

\bibitem{Joe}
J. Fitzsimons, J. Jones, and V. Vedral,
arXiv:1302.2731


\bibitem{Oreshkov}
O. Oreshkov, F. Costa, and C. Brukner, 
Nat. Comm. {\bf3}, 1092 (2012).

\bibitem{Brukner1}
C. Brukner, arXiv:1404.0721

\bibitem{Brukner2}
C. Brukner, Nat. Phys. {\bf10}, 259 (2014).

\bibitem{SWolf}
A. Baumeler, A. Feix, and S. Wolf, arXiv:1403.7333

\bibitem{SWolf2}
A. Baumeler and S. Wolf, arXiv:1312.5916

\bibitem{MorimaePM}
T. Morimae, Phys. Rev. A {\bf90}, 010101(R) (2014).

\bibitem{Hardy1}
L. Hardy, arXiv:0509120

\bibitem{Hardy2}
L. Hardy, 
J. Phys. A {\bf40}, 3081 (2007).

\bibitem{Hardy3}
L. Hardy,
arXiv:0701019


\bibitem{Vedrans}
We call this notation Vedran's notation.

\bibitem{Giulio_note}
G. Chiribella, G. M. D'Ariano, and P. Perinotti,
Euro. Phys. Lett. {\bf83}, 30004 (2008).
\bibitem{Giulio_note2}
G. Chiribella, G. M. D'Ariano, and P. Perinotti,
Phys. Rev. A {\bf80}, 022339 (2009).

\end{thebibliography}
\end{document}